\documentclass[conference]{IEEEtran}
\IEEEoverridecommandlockouts
\usepackage{cite}
\usepackage{amsmath,amssymb,amsfonts}
\usepackage{graphicx}
\usepackage{textcomp}
\usepackage{xcolor}
\usepackage{algorithm}
\usepackage[utf8]{inputenc}
\usepackage{algorithmic}
\ifCLASSOPTIONcompsoc
  \usepackage[nocompress]{cite}
\else
  \usepackage{cite}
\fi
\def\BibTeX{{\rm B\kern-.05em{\sc i\kern-.025em b}\kern-.08em
    T\kern-.1667em\lower.7ex\hbox{E}\kern-.125emX}}
\begin{document}

\title{ CoVid-19 Detection leveraging Vision Transformers and Explainable AI \\}

\author{\IEEEauthorblockN{Pangoth Santhosh Kumar}
\IEEEauthorblockA{\textit{Data Science and Artificial Intelligence} \\
\textit{IIIT Naya Raipur}\\
pangoth20102@iiitnr.edu.in}
\and
\IEEEauthorblockN{Kundrapu Supriya}
\IEEEauthorblockA{\textit{Computer Science and Engineering} \\
\textit{IIIT Naya Raipur}\\
kundrapu20100@iiitnr.edu.in}
\and
\IEEEauthorblockN{Mallikharjuna Rao K}
\IEEEauthorblockA{\textit{Data Science and Artificial Intelligence} \\
\textit{IIIT Naya Raipur}\\
mallikharjuna@iiitnr.edu.in}
\and
\IEEEauthorblockN{Taraka Satya Krishna Teja Malisetti }
\IEEEauthorblockA{\textit{IT Architect/ DevOps Manager} \\
\textit{BMW Group Information Technology Research Center}\\
tarakteja.malisetti@gmail.com}
}

\author{\IEEEauthorblockN{
\large Pangoth Santhosh Kumar\IEEEauthorrefmark{1}, Kundrapu Supriya \IEEEauthorrefmark{1}, Mallikharjuna Rao K\IEEEauthorrefmark{1}, Taraka Satya Krishna Teja Malisetti\IEEEauthorrefmark{2}\\}
\IEEEauthorblockA{\IEEEauthorrefmark{1}\normalsize Department of Computer Science, IIIT Naya Raipur, India}
\IEEEauthorblockA{\IEEEauthorrefmark{2}\normalsize Department of IT Architect/ DevOps Manager, BMW Group Information Technology Research Center}
\normalsize {e-mail:  pangoth20102@iiitnr.edu.in, kundrapu20100@iiitnr.edu.in, mallikharjuna@iiitnr.edu.in, tarakteja.malisetti@gmail.com}
}

\maketitle

\begin{abstract}
Lung disease is a common health problem in many parts of the world.
It is a significant risk to people's health and quality of life all across the globe since it is responsible for five of the top thirty leading causes of death. Among them are COVID-19, pneumonia, and tuberculosis, to name just a few. It is critical to diagnose lung diseases in their early stages. Several different models including machine learning and image processing have been developed for this purpose. The earlier a condition is diagnosed, the better the patient's chances of making a full recovery and surviving into the long term. Thanks to deep learning algorithms, there is significant promise for the autonomous, rapid, and accurate identification of lung diseases based on medical imaging. Several different deep learning strategies, including convolutional neural networks (CNN), vanilla neural networks, visual geometry group-based networks (VGG), and capsule networks, are used for the goal of making lung disease forecasts. The standard CNN has a poor performance when dealing with rotated, tilted, or other aberrant picture orientations. As a result of this, within the scope of this study, we have suggested a vision-transformer-based approach end to end framework for the diagnosis of lung disorders. In the architecture, data augmentation, training of the suggested models, and evaluation of the models are all included. For the purpose of detecting lung diseases such as pneumonia, Covid-19, lung opacity, and others, a specialised Compact Convolution Transformers (CCT) model have been tested and evaluated on datasets such as the Covid-19 Radiography Database. The model has achieved a better accuracy for both its training and validation purposes on the Covid-19 Radiography Database .  There have been a number of different evaluation criteria utilised, such as accuracy, recall, confusion matrix, and so on. been used for the purpose of analysing the model. And than these we also used XAI for evaluationg the model.
\end{abstract}

\section{Introduction}  
Infections of the lungs are persistent illnesses that have an effect on the human body's tissues and organs and make it difficult to breathe.A few examples of lung diseases include pneumonia, Covid-19, tuberculosis, lung cancer, and other other lung problems. According to the Forum of International Respiratory Societies \cite{article5}, approximately 334 million people around the world suffer from asthma. Additionally, each year 1.4 million people pass away as a result of tuberculosis, 1.6 million pass away as a result of lung cancer, and millions more pass away as a result of pneumonia. The COVID-19 pandemic resulted in the infection of millions of individuals, which in turn had a detrimental impact on the healthcare systems located across \cite{DBLP:journals/corr/abs-1711-05225}. There is no question that disorders affecting the lungs are among the leading causes of death and disability across the globe. Early detection considerably boosts both the patient's chances of making a full recovery and their chances of surviving for a longer period of time \cite{Wang_2020}. Healthcare informatics is the field that deals with the management and use of information in the healthcare industry. One example of how machine learning and deep learning algorithms can be used in healthcare informatics is in the detection of lung diseases, such as pneumonia. The widespread distribution of COVID-19 has increased interest in the early detection of lung diseases, as the virus can cause severe lung damage and respiratory issues  \cite{MONDAL2020100374}. In addition, other viral or bacterial infections can also contribute to pneumonia, a type of lung disease. Two commonly used diagnostic methods for finding lung disorders are computed tomography (CT) scans and CXR imaging \cite{inproceedings}. CT scans use X-rays to produce detailed images of the inside of the body, while CXR imaging uses a special camera to produce images of the chest. These diagnostic methods can be used to identify abnormalities in the lungs that may indicate the presence of a lung disease.

\begin{figure}[ht]
    \centering
    \includegraphics[width=\linewidth]{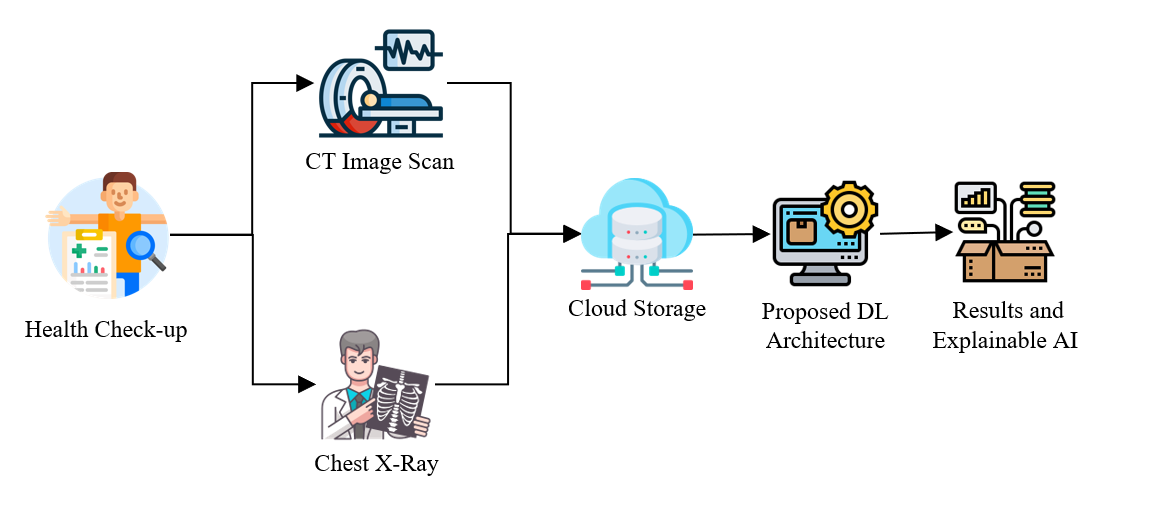}
    \caption{The functionality of the proposed solution}
    \label{fig:overview}
\end{figure}

On the other hand, medical imaging treatments like X-ray and CT-based screening are more often used since they are more readily accessible, they are quicker, and they are typically safe. Imaging using X-rays, as opposed to imaging with CT, is often used for COVID-19 screening since it needs less imaging time and is less expensive than CT imaging. Scanners that use X-ray technology are frequently available even in more remote locations. In recent years, academics have focused their attention on developing methods for diagnosing lung diseases using a variety of methods, including classical machine learning and deep learning. CNN, VGGNet, ResNet, and LSTM are among the many algorithms that may be used in the process of diagnosing lung disorders. In this study, we suggest a transformer-based architecture for the classification and diagnosis of lung diseases. The following is a list of the significant contributions that this work makes:
\begin{itemize}
    \item The use of CT Scan and CXR datasets for the purpose of training the transformer models.
     \item Constructed and analysed the sophisticated image classification model with transformers for the identification of lung disease.
     \item Conducting an analysis of the trained models by employing a variety of accuracy measures, such as precision and recall, amongst others.
\end{itemize}

\section{Related Works}
Several computer-aided design (CAD) systems have been developed throughout the years to assist physicians in deciphering medical images \cite{974918}. However, developing a reliable CAD system is challenging. More choices in how to build such a system have emerged because to the development of powerful graphics processing units (GPUs) and deep learning approaches like convolutional neural networks (CNNs). Chest X-rays, CT scans, histopathology images, etc. may all be used in conjunction with deep learning methods to diagnose lung disorders. Briefly summarise the latest studies that have been undertaken by various researchers to diagnose lung disease utilising chest x-ray and CT scan pictures in the next two subsections.

\subsection{ X-Ray Imaging for the Diagnosis of Lung Disease}

X-ray pictures provide a number of challenges for clinical interpretation, including various possible abnormalities and complicated backgrounds\cite{CHEN2019101554}.
This calls for expert-level manual annotation (radiologists). The ability to automatically analyse X-ray images is rapidly becoming into an important diagnostic resource.
Since recent advancements in this area \cite{MOUSAVI202263} have made deep neural networks widely used, they are often used to the problem of X-ray picture categorization. When compared to other pneumonias, COVID-Net \cite{article} is the only open-source, created, and maintained technology that can reliably identify between COVID-19 and other pneumonias. From the provided specs and preliminary design prototype, COVID-Net is able to learn the architectural design starting point and go further with machine-driven design exploration. Your chest X-ray will work with it. Classification as "normal," "pneumonia," or "COVID-19." Nahid et al \cite{s20123482} .'s method varied the gathering of discriminative features by fusing image processing techniques with a two-channel CNN. The model correctly identified cases of pneumonia with a sensitivity of 97.92\%. If done by hand, however, the crop operation needed to remove the unwanted components would be time-consuming and difficult. In \cite{article2}, a CNN model that makes use of GradCAM to draw attention to affected regions is shown. The model has a good validation accuracy of 84.8\%. However, the authors did not use any augmentation methods to produce unique training samples; as a consequence, the samples were identical. In order to construct models for medical imaging, transfer learning is often used as a workaround for the scarcity of available training data. Chouhan et al. \cite{app10020559} employed a transfer learning technique to classify X-rays of pneumonia.
They pooled the knowledge of five trained models to make pneumonia diagnosis more accurate. To diagnose pneumonia, Rahman et al. \cite{Rahman_2020} used a transfer learning strategy based on convolutional neural networks. We employed models such as AlexNet, ResNet18, DenseNet201, and SqueezeNet to identify chest X-rays that showed germs, viruses, or were just normal. The accuracy rate after using their approach was 98
\subsection{ CT scan for the Diagnosis of Lung Disease}
Roy, Sirohi, and Patle \cite{article1} developed a method for detecting lung cancer nodules using a due to uncontrollable system as well as an adaptive thresholding model. Using grey transformation, this method boosts contrast in the visual field. Segmentation is performed using an active contour model once an image has been binarized. The diagnosis of cancer is categorised using fuzzy inference. Features are retrieved for use in training the classifier, and they include the area, mean, entropy, correlation, main axis length, and minor axis length. The total accuracy of the system is 94.12\%. Given its constraints, the suggested model cannot be used to differentiate between benign and malignant tumours.
The K mean unsupervised learning method is used by Sangamithra and Govindaraju \cite{7566533} for classification or clustering purposes.
It classifies the dataset of pixels according to several characteristics. In order to classify data, this model employs a back propagation neural network.
Features including entropy, correlation, homogeneity, PSNR, and SSIM may be retrieved via the gray-level co-occurrence matrix (GLCM) method. Roughly 90.7\% accuracy may be expected from the system. Using a median filter, which is often used for picture improvement, may aid our new model in filtering out noise and improving accuracy.

\subsection{Algorithm}

\begin{algorithm}

\caption{COVID-19 Detection using Vision Transformers}
\label{alg:covid_detection}
\begin{algorithmic}[1]
\STATE \textbf{Input}: Dataset of CXR = $\{I_1, I_2, \ldots, I_n\}$
\STATE  Load pretrained $\texttt{ViT}$ Model for COVID-19 detection 
\STATE  Load and preprocess the scan report dataset
\STATE Set $Th \xleftarrow{} \text{Threshold}$
\STATE Initialize $Pr \xleftarrow{} \text{COVID19 Positive Report}$
\STATE Initialize $Nr \xleftarrow{} \text{COVID19 Negative Report}$
\FOR{$i = 1$ \TO $n$}
    \STATE Generate features: $F_i = \texttt{ViT}(I_i)$
    \STATE $\text{Prediction}_i = \text{Model}(F_i)$
    \STATE $\text{Confidence}_i = \text{Confidence\_Score}(\text{Prediction}_i)$
    \IF{$\text{Confidence}_i > \text{Threshold}$}
        \STATE $I_i \xrightarrow{ADD} {Pr}$ 
    \ELSE
         \STATE $I_i \xrightarrow{ADD} {Nr}$ 
    \ENDIF
\ENDFOR
\STATE \textbf{Output} $Pr,Nr$
\end{algorithmic}
\end{algorithm}
The algorithm introduces a methodical process for COVID-19 detection using Vision Transformers on scan report images, showcasing the integration of advanced image analysis techniques into medical diagnostics. It commences with dataset initialization, employing a pretrained Vision Transformer ($\texttt{ViT}$) model renowned for capturing image intricacies. Subsequent preprocessing readies the dataset for compatibility with the model. A threshold ($Th$) is set as a confidence benchmark for negative COVID-19 detection, segregating reports into $Pr$ (COVID19 Positive Reports) and $Nr$ (COVID19 Negative Reports) lists. Within the scan report loop, $\texttt{ViT}$ model-derived features ($F_i$) enter a classification model ($\text{Model}$) for COVID-19 prediction, with resulting confidence ($\text{Confidence}_i$) aiding categorization. Exceeding the threshold marks reports as COVID-19 positive, aggregating in $Pr$, while those below join $Nr$. Concluding, the algorithm furnishes $Pr$ and $Nr$, presenting a potent data-driven approach merging Vision Transformers and predictive modeling for COVID-19 diagnosis support in medical practice.

\section{Research Methodology}
In this subsection, we will talk about our suggested unique architecture for defect diagnosis of X-Ray and CT Scan pictures. This design primarily consists of three phases: image pre-processing, image augmentation, and our Vision Transformers architecture for automatically extracting ROIs, classifying images and identification of diseases.

\begin{figure}[ht]
    \centering
    \includegraphics[width=\linewidth]{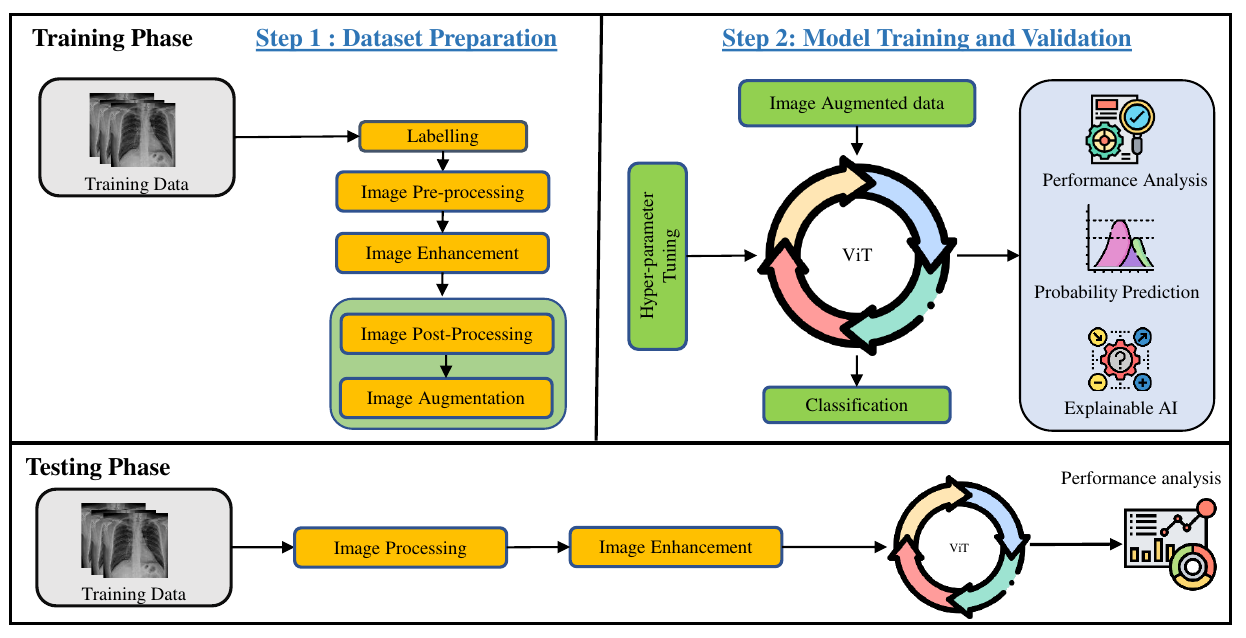}
    \caption{The Overview of Proposed Solution}
    \label{fig:overview}
\end{figure}

\begin{figure*}[ht]
    \centering
    \centerline{\includegraphics[height=37mm,width=175mm]{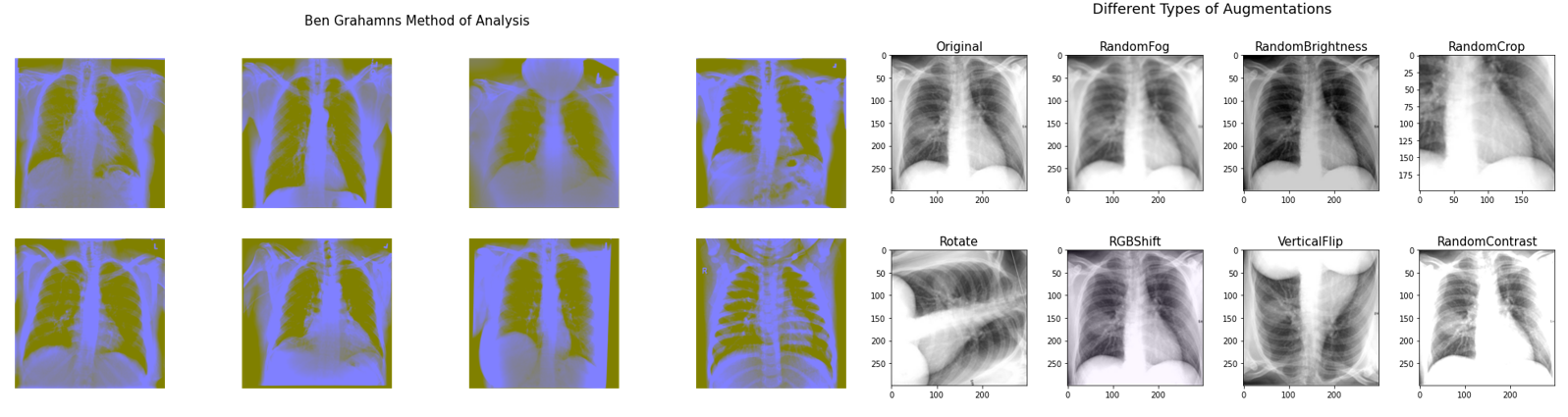}}
    \caption{The Ben Graham Processing and Data Augmentation of the images}
    \label{img_4}
\end{figure*} 

\subsection{Contrast limited adaptive histogram equalization}\label{AA}
X-Ray and CT-Scan photographs used in the medical field often include noise from the medical field, such as noise in the backdrop that is hazy. The poor performance of the prediction models may be attributed to the presence of  noise in an X-Ray picture. Aside from the background noise, it is vital to extract the area of interest (ROI) in order to improve the performance of predictive algorithms, hence minimizing the amount of redundant information and the amount of work required for computational pattern recognition. The vast majority of the present architectural designs do not take these critical aspects into account; instead, they force unfiltered imagery to fit into a predicted framework. Taking all of these considerations into account, the approach that has been suggested improves the quality of the picture and derives the ROI by contrast-limited adaptive histogram equalization (CLAHE). This strategy, in contrast to the designs that are now in use, lessens the amount of computing work required and boosts the predictive models' overall performance. The issue of contrast over-amplification is addressed by CLAHE, which is a modification of the technique known as adaptive histogram equalization (AHE). CLAHE works with discrete parts of a picture, which are referred to as tiles, rather than analyzing the whole image. Tiles that are next to one another are blended together using bilinear interpolation in order to get rid of arbitrary boundaries.

\subsection{Ben Graham Method: }
Contrast CLAHE, or limited adaptive histogram equalization, is used with Ben Graham's preprocessing approach, which involves eliminating the colour that is considered to be the local average. When compared to utilizing either of the algorithms on its own, this results in an improvement in the clarity of the vessels in the majority of the photos. However, after conducting a number of trials, it was shown that the performance of the combination of CLAHE and Ben Graham's method is marginally worse than that of CLAHE when applied to certain circumstances, particularly the Chase dataset. In order to address this issue, a feature fusion method is used, which consists of concatenating the solitary CLAHE preprocessed picture with the combined CLAHE and Ben Graham's preprocessed image. This method ranks highest among all of the tests that were carried out on all three datasets.

\subsection{Data Augmentation}
Because more convolutional networks do not have inductive biases, ViT has only been validated as a benchmark design if it is trained on big datasets. This is because more convolutional networks do not have inductive biases. In this study, we adapted the idea from the analysis presented by Steiner et al., and the approach we suggested provides image augmentation to ViT-based designs in a cautious manner, resulting in higher efficiency. In addition to this, the relevance of ViT frameworks may be increased by the visual alternatives that are provided through augmentation. As a result, the pictures of the ROI that were extracted are subjected to further processing, during which they are altered visually in a variety of ways using image enhancement methods such as gaussian blur, random rotation, zooming, and flipping in a variety of orientations.

\subsection{Compact Convolution Transformers}

The convolutional block in CCT allows the model to capture spatial relationships between patches of input data, which can be useful for tasks such as image classification. This is because a convolutional layer is able to learn local patterns in the data and is able to take into account the spatial structure of the input. By using a max pooling layer after the convolution, the model is able to reduce the dimensionality of the output while maintaining the most important information. The ReLU activation function helps the model learn non-linear relationships between the input and the output. Overall, the use of the convolutional block in CCT can improve the model's performance on certain tasks compared to the patch and embedding approach used in ViT.

The convolutional block in CCT is applied to the feature map of an input image. The feature map, represented by L, is a representation of the input image that encodes important information such as edges and textures. In the convolutional block, the feature map is first passed through a Conv2d operation with D filters, where D is the embedding dimension of the transformer backbone. This operation applies a set of D filters to the feature map, each of which is able to learn a different pattern in the data. Next, the output of the Conv2d operation is passed through a max pooling layer. This layer takes the maximum value of each filter output, which helps reduce the dimensionality of the output while preserving the most important information. The output of the max pooling layer is then passed through a ReLU activation function, which helps the model learn non-linear relationships between the input and the output.

 \begin{equation}
    Image (L) \in  \mathbb{R}^{H\times W\times C}
    \label{eq3}
\end{equation}

 \begin{equation}
    L_{0} = MaxPool(ReLU(Conv2D(L)) 
    \label{eq3}
\end{equation}

Overall, the use of the convolutional block in CCT allows the model to capture spatial relationships between patches of input data and improve its performance on tasks such as image classification. This is because the convolutional block is able to learn local patterns in the data and take into account the spatial structure of the input, which is useful for these types of tasks.

Incorporating a convolutional phase in CCT allows the technique to be more flexible and adaptable than techniques like ViT. This is because the convolutional phase allows the model to process input images of any resolution, rather than being limited to input images that are exactly divided by the patch size, as is the case with ViT. The use of convolutional blocks in CCT is also more effective in producing tokens for the transformer. Because convolutional blocks are able to learn local patterns in the data, they are able to generate tokens that better capture the information in the input image. This can be useful for tasks such as image classification, where the model needs to be able to extract important features from the input image in order to make accurate predictions. Furthermore, the convolutional blocks in CCT can be repeated for additional downsampling, which allows the model to learn hierarchical representations of the input data. This can improve the model's performance on certain tasks, as it allows the model to learn both local and global patterns in the data. Additionally, the number of convolutional blocks and the downsampling ratio can be adjusted to suit the specific needs of the task at hand.
\begin{figure*}[ht]
    \centering
    \centerline{\includegraphics[width=\linewidth]{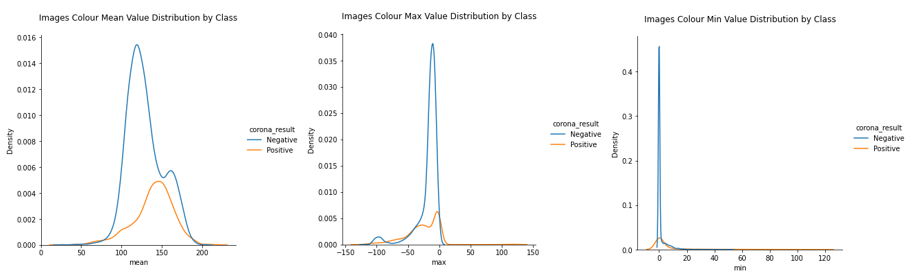}}
    \caption{Data Distribution with respect to mean, max and min value}
    \label{img_4}
\end{figure*} 
In ViT, the transformer is used with class tokenization, which involves dividing the input data into patches and treating each patch as a separate token \cite{medi2023novel}. In CCT, the transformer is used with sequence pooling, which is an attention-based technique that pools information from the entire output token sequence. This is different from class tokenization, as it retains information from the entire sequence rather than just individual patches. The use of sequence pooling in CCT is motivated by the fact that the output token sequence contains important information that spans multiple regions of the input data. Retaining this information can improve the model's performance on certain tasks. Additionally, using sequence pooling reduces the number of tokens that need to be processed by the transformer, which can slightly reduce the computation required for the model. Overall, the use of sequence pooling in CCT can improve the model's performance and make it more efficient compared to using class tokenization in ViT as represented in \cite{krishna2022epersist}. The output sequence is mapped in this operation using the transformation $ T: \mathbb{R}^{B\times L\times D} $
\begin{equation}
Z_{S} = f(L_{0})\in \mathbb{R}^{B\times L\times D}
\end{equation}

In CCT, the output of the transformer encoder, represented by $Z_{S}$, is passed through a linear layer $g(Z_{S})$. The linear layer, represented by $g(Z_{S})$, maps the output of the transformer encoder to a vector of size $D \times 1$, where D is the total embedded dimension of the input data. The transformer encoder architecture of CCT is inspired by \cite{krishna2023lesionaid}. This vector is then passed through a softmax activation function, which produces a probability distribution over the possible classes of the input data. The use of the linear layer and the softmax activation function in CCT allows the model to make predictions about the class of the input data. The linear layer maps the output of the transformer encoder to a vector of size $D \times 1$, which encodes important information about the input data. The softmax activation function then converts this vector into a probability distribution over the possible classes, which allows the model to make a prediction about the class of the input data. Overall, this process allows CCT to effectively classify input data.
\begin{equation}
    {Z}^{`}_{S}  = softmax({g(Z_{S})^T}) \in \mathbb{R}^{D\times1\times n} 
\end{equation}

\begin{equation}
   Z = {Z_{S}} {Z}^{`}_{S}  = softmax({g(Z_{S})^T}) \times {Z_{S}} \in \mathbb{R}^{B\times1\times D} 
\end{equation}
Equation 10 in CCT generates an output for each input token, which is then processed by equation 11. By flattening the output, $Z \in \mathbb{R}^{B \times 1 \times D}$, the model produces a representation of the input data that is suitable for input to the classifier. The use of sequence pooling in CCT allows the network to associate data from throughout the input information and evaluate the sequential embeddings of the latent space created by the transformer encoder. This allows the model to capture important information from the entire input sequence, rather than just individual patches, which can improve its performance on certain tasks. Overall, CCT is a model that incorporates a convolutional tokenizer, sequence pooling, and a transformer encoder. This combination of techniques allows CCT to effectively classify input data by capturing spatial relationships and important information from the entire input sequence.
\section{Experimental Results and Analysis}
\subsection{Dataset Preperation}
For the purpose of training the ViT model, this experiment makes use of a sleepiness estimate benchmark dataset called COVID-19 Radiography Database. This database is one of the most consid- erable datasets accessible among the public datasets that are currently available. The COVID-19 Radiography Training Database has a total of 36 participants, each with their own unique scenario. We gather random photos of each participant and identify them using a binary system according to whether or not they have covid. This dataset contains frames with a resolution of 640 by 480, which is considered to be high in comparison to the resolution of other COVID19 datasets. Significant adjustments have been made to the size, position, and impressions of the chests included in the dataset.
As a result, this benchmark dataset is suitable for demonstrating the efficacy and performance in real-world contexts.
\subsection{Experimental Setup and Computational Specification}

This section covers the hardware and software computing requirements for the proposed system. The framework was developed using the Tensorflow 2.0, and OpenCV libraries and was implemented in Python 3.9. The minimum requirements for the proposed system are listed in the table.

\begin{table}[ht]
\caption{Minimum Requirement for the Proposed System}
\label{table1}
\begin{center}
{
\begin{tabular}{|c|c|}
\hline
\textbf{System Specifications} & \textbf{Configuration Details}  \\ 
\hline
Operating System & Windows 11\\
\hline
Random Access Memory &  32GB \\
\hline
System Type & 64 bit Windows OS\\
\hline
CPU & Intel core i9\\
\hline
GPU & GeForce 3070 Ti\\
\hline
Clock Speed & 4.80Ghz\\
\hline
Python & 3.10.4\\
\hline
Frameworks& TensorFlow, Keras\\
\hline
\end{tabular}
}
\end{center}
\end{table}

\begin{figure*}[ht]
    \centering
    \centerline{\includegraphics[width=\linewidth]{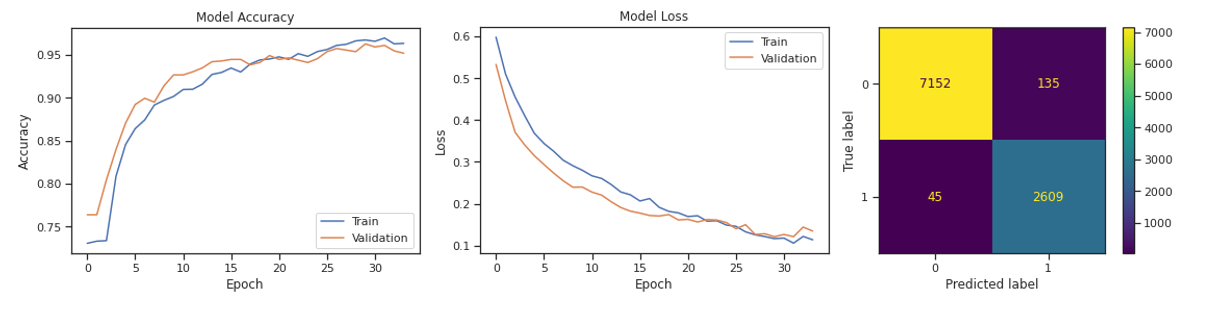}}
    \caption{Learning Curves of Training and Validation accuracy and Loss for 150 Epochs and its Confusion Matrix}
    \label{img_4}
\end{figure*} 

\subsection{Distribution of Data}
The Distribution of the colour values in the images are taken into consideration based on the target class to extract useful insights. 
Insights for pixels from the plot of Mean vs Density:
\begin{itemize}
    \item Covid Negative situations have a maximum pixel value that is more than 0.014 but less than 0.016.
    \item Greater than 0.004 but less than 0.006 is the maximum value for a pixel in circumstances when Covid is positive.
\end{itemize}

The Max versus Density plot reveals the following observations about pixels:
\begin{itemize}
    \item For Covid Negative situations, the maximum pixel value must be larger than 0.035 and less than 0.040.
    \item The maximum pixel value for situations that are considered to be Covid Positive is 0.005.
\end{itemize}

The Minimum versus Density graphic reveals the following about pixels:
\begin{itemize}
    \item The maximum pixel value for situations that are classified as Covid Negative is more than 0.4.
    \item In circumstances when Covid is positive, the maximum pixel value must be more than 0.0 and less than 0.1.

\end{itemize}

\subsection{Evaluating Compact Convolutional Vision Transformers}
For the goal of conducting an analysis of the effectiveness of our ViT system, conventional benchmarks for judging classification models have been used. The learning curves of accuracy and loss that were experienced throughout the training and validation of the models are shown in the figure below. Because both the validation curve and the training curve maintain a point of stability with a minimal variation between them, these learning plots provide indicative of an efficient learning method. The training of the efficient ViT model was developed to incorporate three distinct but interrelated tasks at the same time. These tasks are as follows: 1) the calculation of output; 2) the correction of mistakes, and 3) the fine-tuning of the hyper-parameters. The purpose of this design was to achieve the best possible results from the training of the model. When employing a certain combination of hyper-parameters, the maximum training and validation accuracy, respectively, were determined to be 97\% when using a particular combination of hyper-parameters. This was discovered after a number of rounds during which the hyper-parameters were fine-tuned.

In order to conduct further assessments of the efficacy of the classification ViT model, calculations of hamming loss and binary cross-entropy are carried out. It has been determined that the cross-entropy loss and the hamming loss for the trained ViT model both equal 0.6907 and that the corresponding hamming loss and cross-entropy losses are 0.0673. The fact that the log loss was coming closer and closer to zero was a sign that beneficial results were being achieved. Cross entropy loss had disproportionately penalized erroneous predictions, which is a factor that is significant for a loss function but undesirable for a metric. This is because this factor is a factor that is important for a loss function but bad for a metric. Cross entropy loss was the method that was used. As a result, scores for precision, recall, and F1, in addition to those for several other forms of accuracy, were calculated and shown.

\begin{table}[ht]
\caption{Calculated Evaluation metrics}
\label{table6}
\begin{center}
\scalebox{0.9}
{
\begin{tabular}{|c|c|c|c|c|}
\hline
 & Precision & Recall & F1 Score & Image Support \\ 
\hline
Healthy & 0.97 & 0.98 & 0.97 & 844\\
\hline
Diseased & 0.98 & 0.97 & 0.95 & 261\\
\hline
Macro Average & 0.93 & 0.94 & 0.93 &  1105\\
\hline
Weighted Average & 0.95 & 0.95 & 0.95 &  1005\\
\hline
\end{tabular}
}
\end{center}
\end{table}

\subsection{Grad-CAM: Gradient-weighted Class Activation Mapping}
Gradient-weighted Class Activation Mapping (Grad-CAM) is a technique used in deep learning to generate a heatmap highlighting the regions in an image that are most important for a specific prediction. This can be useful for understanding how a model is making a particular decision, and can also help with debugging and improving the model.

\begin{figure}[ht]
    \centering
    \includegraphics[width=\linewidth]{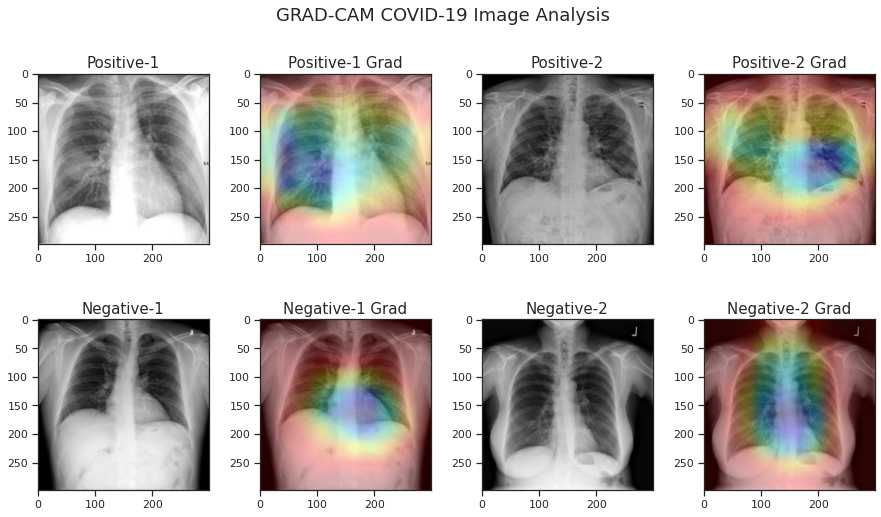}
    \caption{Activation Maps of the Sample Predictions}
    \label{fig:overview}
\end{figure}

The following are the insights obtained by the XAI :
\begin{itemize}
    \item \textbf{Positive-1:} In its Grad-CAM picture, on the right midsection of it we can see the blue colour highlighted piece, which is opacity owing to which it belongs to COVID's Positive Category. Because of this, it is a sample that has been given the Positive-1 designation.

    \item \textbf{Positive-2:} In its Grad-CAM picture, on the left bottom half of it we can see the blue green hue highlighted region which is consolidation. Because there is no Tree-Bud owing to this, it is classified as a member of the COVID - Positive Category.

    \textbf{Negative-1:} In its Grad-CAM picture, we are able to view the blue colour highlighted section that is between the Cardiac and the Diaphragm. Since there was no opacity identified, it falls into the COVID - Negative Category because of this.

    \item \textbf{Negative-2:} Within its Grad-CAM picture, we are able to view the blue colour part that emphasizes the Trachea. As a result of the absence of any other opacity, it is classified as falling within the COVID - Negative Category.
\end{itemize}

\section{Conclusion}
We have developed Compact Convolutional Transformers for lung disease classification in this work. Apart from this we have developed an end-to-end framework which consists of three phases for classification, phase 1 consists of image preprocessing using contrast limited adaptive histogram equalization (CLAHE) \& ben graham method, phase 2 includes image augmentation and finally phase 3 consists of the compact convolution transformer for classification. CLAHE is operated on small regions of an image and is used to improve the visibility level of a foggy image and ben graham method is used to eliminate the color that is considered to be the local average. And in the phase 2 we have used different data augmentation techniques such as rotation, zooming, flipping, etc. And finally the CCT model is trained using the training dataset and it has obtained 97\% and 94.6\% as training and validation accuracy respectively. Explainable AI has been implemented in order to comprehend the reasoning behind a specific decision or action, or to provide an understanding of how the AI system operates. GradCAM is a technique used to visualize the region of an input that is used to predict the lesion with the ViT model. 

\bibliography{dip.bib}

\begin{thebibliography}{10}

\bibitem{article5}
Mona Hellou, Fulvia Mazzaferri, Eleonora Cremonini, Elisa Gentilotti, Pasquale
  De~Nardo, Itamar Poran, Mariska Leeflang, Evelina Tacconelli, Mical Paul, and
  Anna Górska.
\newblock Systematic review nucleic acid amplification tests on respiratory
  samples for the diagnosis of coronavirus infections: a systematic review and
  meta- analysis.
\newblock {\em Clinical Microbiology and Infection}, 27, 11 2020.

\bibitem{DBLP:journals/corr/abs-1711-05225}
Pranav Rajpurkar, Jeremy Irvin, Kaylie Zhu, Brandon Yang, Hershel Mehta, Tony
  Duan, Daisy~Yi Ding, Aarti Bagul, Curtis~P. Langlotz, Katie~S. Shpanskaya,
  Matthew~P. Lungren, and Andrew~Y. Ng.
\newblock Chexnet: Radiologist-level pneumonia detection on chest x-rays with
  deep learning.
\newblock {\em CoRR}, abs/1711.05225, 2017.

\bibitem{Wang_2020}
Zhao Wang, Quande Liu, and Qi~Dou.
\newblock Contrastive cross-site learning with redesigned net for {COVID}-19
  {CT} classification.
\newblock {\em {IEEE} Journal of Biomedical and Health Informatics},
  24(10):2806--2813, oct 2020.

\bibitem{MONDAL2020100374}
M.~Rubaiyat~Hossain Mondal, Subrato Bharati, Prajoy Podder, and Priya Podder.
\newblock Data analytics for novel coronavirus disease.
\newblock {\em Informatics in Medicine Unlocked}, 20:100374, 2020.

\bibitem{inproceedings}
Arata Saraiva, N.~Ferreira, Luciano Sousa, Nator Carvalho~da Costa, José
  Sousa, D.~Santos, Antonio Valente, and Salviano Soares.
\newblock Classiﬁcation of images of childhood pneumonia using convolutional
  neural networks.
\newblock pages 112--119, 01 2019.

\bibitem{974918}
B.~Van~Ginneken, B.M. Ter Haar~Romeny, and M.A. Viergever.
\newblock Computer-aided diagnosis in chest radiography: a survey.
\newblock {\em IEEE Transactions on Medical Imaging}, 20(12):1228--1241, 2001.

\bibitem{CHEN2019101554}
Bingzhi Chen, Jinxing Li, Xiaobao Guo, and Guangming Lu.
\newblock Dualchexnet: dual asymmetric feature learning for thoracic disease
  classification in chest x-rays.
\newblock {\em Biomedical Signal Processing and Control}, 53:101554, 2019.

\bibitem{MOUSAVI202263}
Zohreh Mousavi, Nahal Shahini, Sobhan Sheykhivand, Sina Mojtahedi, and Afrooz
  Arshadi.
\newblock Covid-19 detection using chest x-ray images based on a developed deep
  neural network.
\newblock {\em SLAS Technology}, 27(1):63--75, 2022.

\bibitem{article}
Linda Wang, Zhong Lin, and Alexander Wong.
\newblock Covid-net: a tailored deep convolutional neural network design for
  detection of covid-19 cases from chest x-ray images.
\newblock {\em Scientific Reports}, 10, 11 2020.

\bibitem{s20123482}
Abdullah-Al Nahid, Niloy Sikder, Anupam~Kumar Bairagi, Md.~Abdur Razzaque,
  Mehedi Masud, Abbas Z.~Kouzani, and M.~A.~Parvez Mahmud.
\newblock A novel method to identify pneumonia through analyzing chest
  radiographs employing a multichannel convolutional neural network.
\newblock {\em Sensors}, 20(12), 2020.

\bibitem{article2}
Hai Nguyen, Hoang Huynh, Toan Tran, and Hiep Huynh.
\newblock Explanation of the convolutional neural network classifying chest
  x-ray images supporting pneumonia diagnosis.
\newblock {\em EAI Endorsed Transactions on Context-aware Systems and
  Applications}, 7:165349, 07 2018.

\bibitem{app10020559}
Vikash Chouhan, Sanjay~Kumar Singh, Aditya Khamparia, Deepak Gupta, Prayag
  Tiwari, Catarina Moreira, Robertas Damaševičius, and Victor Hugo~C.
  de~Albuquerque.
\newblock A novel transfer learning based approach for pneumonia detection in
  chest x-ray images.
\newblock {\em Applied Sciences}, 10(2), 2020.

\bibitem{Rahman_2020}
Tawsifur Rahman, Muhammad E.~H. Chowdhury, Amith Khandakar, Khandaker~R. Islam,
  Khandaker~F. Islam, Zaid~B. Mahbub, Muhammad~A. Kadir, and Saad Kashem.
\newblock Transfer learning with deep convolutional neural network ({CNN}) for
  pneumonia detection using chest x-ray.
\newblock {\em Applied Sciences}, (9):3233, may 2020.

\bibitem{article1}
Tanushree Roy, Neeraj Sirohi, and Arti Patle.
\newblock Classification of lung image and nodule detection using fuzzy
  inference system.
\newblock {\em International Conference on Computing, Communication and
  Automation, ICCCA 2015}, pages 1204--1207, 07 2015.

\bibitem{7566533}
P.B. Sangamithraa and S.~Govindaraju.
\newblock Lung tumour detection and classification using ek-mean clustering.
\newblock In {\em 2016 International Conference on Wireless Communications,
  Signal Processing and Networking (WiSPNET)}, pages 2201--2206, 2016.

\bibitem{medi2023novel}
Prathistith~Raj Medi, Ghanta~Sai Krishna, Praneeth Nemani, Satyanarayana
  Vollala, and Santosh Kumar.
\newblock A novel end-to-end framework for occluded pixel reconstruction with
  spatio-temporal features for improved person re-identification.
\newblock In {\em 2023 8th International Conference on Business and Industrial
  Research (ICBIR)}, pages 865--870. IEEE, 2023.

\bibitem{krishna2022epersist}
Ghanta~Sai Krishna, Dyavat Sumith, and Garika Akshay.
\newblock Epersist: A two-wheeled self balancing robot using pid controller and
  deep reinforcement learning.
\newblock In {\em 2022 22nd International Conference on Control, Automation and
  Systems (ICCAS)}, pages 1488--1492. IEEE, 2022.

\bibitem{krishna2023lesionaid}
Ghanta~Sai Krishna, Kundrapu Supriya, Meetiksha Sorgile, et~al.
\newblock Lesionaid: Vision transformers-based skin lesion generation and
  classification.
\newblock {\em arXiv preprint arXiv:2302.01104}, 2023.

\end{thebibliography}

\end{document}